\documentclass[aps,prl,twocolumn,noshowpacs,superscriptaddress]{revtex4-1}
\usepackage{graphicx,epsfig, subfigure}
\usepackage{amsbsy}
\usepackage{amssymb}
\usepackage{amsmath}
\usepackage{mathtools}
\usepackage{hyperref}
\usepackage[bottom]{footmisc}
\usepackage{xcolor}
\usepackage{color}
\usepackage{times}
\usepackage{textcomp}
\usepackage{verbatim} 

\date{\today}

\begin{document}

\newcommand{\eqnref}[1]{Eq.~\ref{#1}}
\newcommand{\figref}[2][]{Fig.~\ref{#2}#1}
\newcommand{\RN}[1]{%
  \textup{\uppercase\expandafter{\romannumeral#1}}%
}

%\title{5D cooling and precession-coupled nonlinear dynamics of a levitated nanodumbbell}
\title{5D cooling and nonlinear dynamics of an optically levitated nanodumbbell}

\author{Jaehoon Bang}
	\affiliation{School of Electrical and Computer Engineering, Purdue University, West Lafayette, Indiana 47907, USA}
	\author{Troy Seberson}
\affiliation{Department of Physics and Astronomy, Purdue University, West Lafayette, Indiana 47907, USA}
	\author{Peng Ju}
\affiliation{Department of Physics and Astronomy, Purdue University, West Lafayette, Indiana 47907, USA}
\author{Jonghoon Ahn}
	\affiliation{School of Electrical and Computer Engineering, Purdue University, West Lafayette, Indiana 47907, USA}
\author{Zhujing Xu}
\affiliation{Department of Physics and Astronomy, Purdue University, West Lafayette, Indiana 47907, USA}
\author{Xingyu Gao}
\affiliation{Department of Physics and Astronomy, Purdue University, West Lafayette, Indiana 47907, USA}
\author{Francis Robicheaux}
\affiliation{Department of Physics and Astronomy, Purdue University, West Lafayette, Indiana 47907, USA}
\affiliation{Purdue Quantum Science and Engineering Institute, Purdue University, West Lafayette, Indiana 47907, USA}
		
\author{Tongcang Li}
\email{tcli@purdue.edu}
\affiliation{School of Electrical and Computer Engineering, Purdue University, West Lafayette, Indiana 47907, USA}
\affiliation{Department of Physics and Astronomy, Purdue University, West Lafayette, Indiana 47907, USA}
\affiliation{Purdue Quantum Science and Engineering Institute, Purdue University, West Lafayette, Indiana 47907, USA}
\affiliation{Birck Nanotechnology Center, Purdue University, West Lafayette, Indiana 47907, USA}
	\date{\today}% It is always \today, today,
	% but any date may be explicitly specified

\begin{abstract}
Optically levitated nonspherical particles in vacuum are excellent candidates for torque sensing, rotational quantum mechanics, high-frequency gravitational wave detection, and multiple other applications. Many potential applications, such as detecting the Casimir torque near a birefringent surface, require simultaneous cooling of both the center-of-mass motion and the torsional vibration (or rotation) of a nonspherical nanoparticle.  Here we report the first 5D cooling of a levitated nanoparticle. We cool the 3 center-of-mass motion modes and 2 torsional vibration modes of a levitated nanodumbbell in a linearly-polarized laser simultaneously. The only uncooled rigid-body degree of freedom is the rotation of the nanodumbbell around its long axis. This free rotation mode does not couple to the optical tweezers directly.  Surprisingly, we observe that it  strongly affects the torsional vibrations of the nanodumbbell.  This work deepens our understanding of the nonlinear dynamics and rotation coupling of a levitated nanoparticle and paves the way towards full quantum control of its motion.
 \end{abstract}
\maketitle

In recent years, levitated optomechanics provided a fruitful platform for nonequilibrium thermodynamics \cite{li2010measurement,Gieseler2014,Millen2014,Hoang2018}, nonlinear dynamics \cite{Gieseler2013,Fonseca2016}, precision measurements \cite{Geraci2010ShortRange,Arita2013,DavidMoorePhysRevLett.113.251801,Geraci2016,qvarfort2018gravimetry,ahn2020}, macroscopic quantum mechanics \cite{Chang2009,deli2019motional,Bose2017QuantumGravity}, and several other applications \cite{yin2013optomechanics,millen2020optomechanics}. 
Besides extensive studies on levitated spherical particles, there are growing interests in levitated {\em nonspherical} particles \cite{Romero_Isart_2010,SinghPhysRevLett.105.213602, Bhattacharya_10.1080/09500340.2013.778341,NanodumbbellPhysRevLett.110.143604, Hoang2016,Kuhn2017,SticklerPhysRevLett.121.040401,Rashid2018,Seberson2019,delord2020spin,Ahn2018,laan2020optically}. For example, a levitated nanodumbbell in a linearly-polarized optical tweezer is a novel analogy of the Cavendish torsion balance for precision measurements \cite{Ahn2018}.  With a circularly-polarized laser, it can rotate at record-high GHz frequencies \cite{ahn2020,Ahn2018,laan2020optically,Reimann2018}. Levitated nonspherical particles have also been proposed to  measure the Casimir torque \cite{Xu2017}, create rotational matter-wave interferometers \cite{Stickler2018probing}, and search for high-frequency gravitational waves \cite{Geraci2013}. 

Many potential applications of levitated nonspherical particles require simultaneous cooling of their center-of-mass (c.m.) motions and torsional vibrations (or rotations). For example, to measure the Casimir torque near a birefringent surface with a levitated nanodumbbell or nanorod \cite{Xu2017},  we will need to cool its c.m. motion to prevent loss near the surface, and cool its torsional vibration for detecting the Casimir torque at fixed orientations. To detect the gravitational wave with a levitated microdisk in an optical cavity \cite{Geraci2013}, we will also need to cool its torsional vibrations in addition to c.m. motions to minimize light being scattered out of the optical cavity. Here we report the first 5D (five-degrees-of-freedom) cooling of an optically levitated nanodumbbell. We also investigate the nonlinear dynamics of its motion. Surprisingly, we observe that the free rotation of the nanodummbell around its long axis strongly affects its torsional vibrations. Our work is an essential step towards full quantum control of the rigid-body motion of a levitated nonspherical particle and opens up many potential applications \cite{Xu2017,Stickler2018probing,Geraci2013,Guccione2013mirror}.

\begin{figure}[htb!]
	\includegraphics[scale=0.4]{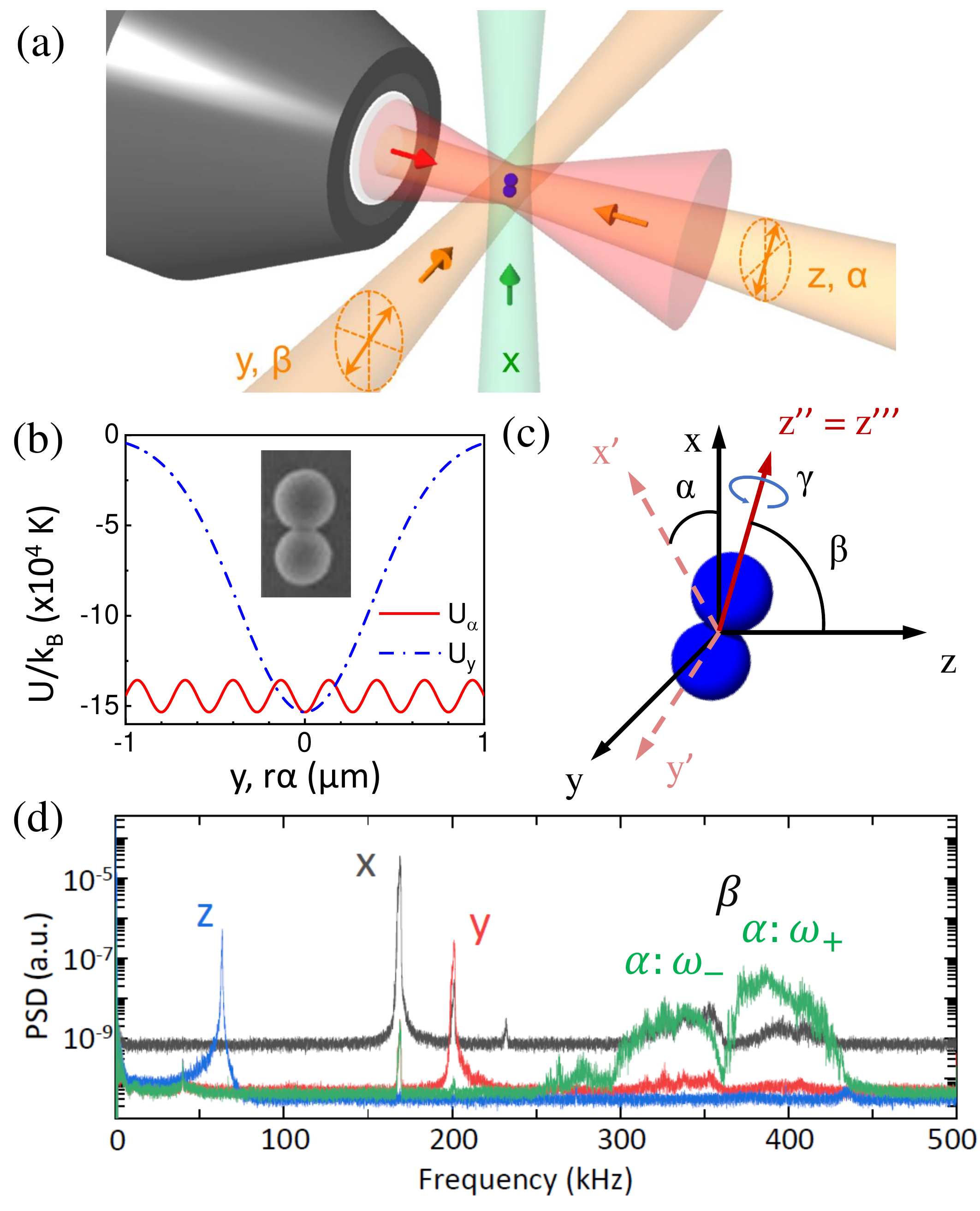}
	\caption{(a) Schematic of 5D cooling of a levitated nanodumbbell. A 1064 nm laser is tightly focused with an NA=0.85 objective lens to levitate a nanodumbbell in vacuum. The 1064 nm trapping laser is polarized in $x$ direction and propagates along $z$-axis as indicated by a red arrow.  Cooling lasers and their directions of propagation are illustrated in green and orange. The polarization directions of the $y$ and $z$ cooling lasers  are tilted to form an angle of about 10 degrees with respect to the polarization direction of the trapping laser. The cooling laser propagating in the $x$ direction is polarized along the $y$ axis. (b) Trapping potential as a function of the c.m. motion along the $y$ axis and the torsional vibration along $\alpha$ direction. For direct comparison, the radius $r=85$~nm  is multiplied to the angle $\alpha$ in the torsional potential case.  Inset is a scanning electron microscope (SEM) image of a nanodumbbell. (c) Definition of the Euler angles $\alpha$, $\beta$, $\gamma$ used in this paper for $z-y'-z''$ convention. For small libration amplitudes, $\alpha$ represents the motion near the $x$-axis in the $x-y$ plane, and $\beta$ represents the motion in the $x-z$ plane. $\gamma$ stands for the free rotation around the $z''=z'''$ axis. $x''$, $y''$, $x'''$, $y'''$ are not shown to simplify the figure. (d) Power spectral density (PSD) measured for a nanodumbbell which consists of two 170 nm-diameter silica spheres. PSDs obtained from four different detectors for $x$, $y$, $z$ and $\alpha$ are shown. Note that the PSD for the $\beta$ motion could be obtained from the high frequency part of the signal obtained by the $x$ detector. These PSDs are taken at a pressure of $3\times10^{-3}$ Torr with $x$, $y$ and $z$ motion cooling. The data acquisition time is 1 sec.}
	\label{scheme}
\end{figure}

In this work, we investigate the torsional nonlinear dynamics of a silica nanodumbbell levitated in a linearly-polarized laser and cool its motion in 5D (Fig.\ref{scheme}). 
 In the focus of a linearly-polarized laser, the long axis of the nanodumbbell tends to align with the polarization direction of the laser. Becasue of the collisions with surrounding air molecules, the nanodumbbell will undergo confined  Brownian motion in 3 translational and 2 torsional vibration modes, and free Brownian rotation around its long axis. The observed nonlinearity in the torsional motion is much stronger than that in its c.m. motion \cite{Gieseler2013}. The strong nonlinearity can potentially be used  for  generating nonclassical  states and sensing \cite{qvarfort2018gravimetry,RevModPhys.86.1391,badzey2005coherent}.
Counter-intuitively, we also find that the thermal Brownian rotation of the nanodumbbell around its long axis strongly affects its torsional vibrations, even though this rotational mode does not couple to the optical tweezers directly. 
After investigating the rigid-body dynamics of a levitated nanodumbbell in all 6 degrees of freedom, we  demonstrate 5D cooling of a levitated nanodumbbell. We cool its c.m. motion to a few K in all three translational directions, and cool its two torsional motions  to about 10 K. 

\begin{figure}[tbh]
	\includegraphics[scale=1]{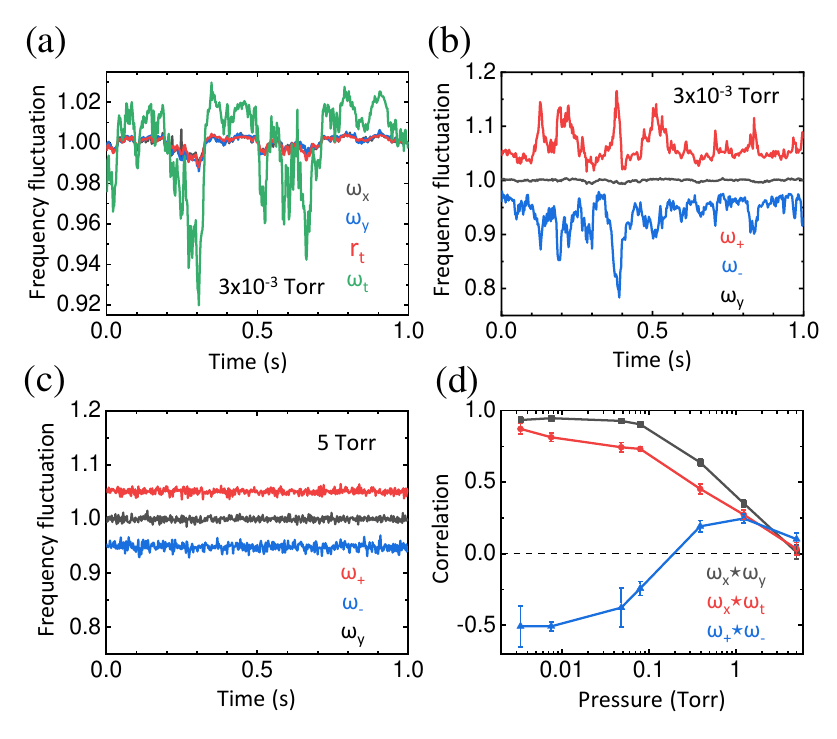}
	\caption{(a) Relative frequency fluctuations for translational motions ($\omega_x$, $\omega_y$) and torsional motions at $3.33\times10^{-3}$ Torr. (b) Relative frequency fluctuations of $\omega_{+}$ (red), $\omega_{-}$ (blue), and $\omega_y$ (grey) at $3.33\times10^{-3}$ Torr. The $\omega_{+}$ and $\omega_{-}$ frequencies are normalized by the average  $\langle(\omega_{+}+\omega_{-})/2\rangle$. $\omega_y$ is normalized by its average $\langle\omega_y \rangle$. (c) Relative frequency fluctuations of $\omega_{+}$ (red), $\omega_{-}$ (blue), $\omega_y$ (grey) at $5$ Torr.   In (a), (b), (c), each frequency is determined from a PSD corresponding to 2 ms of data in time.    (d) Frequency correlations between different degrees of freedom.  The correlations are plotted as a function of pressure. Each data point corresponds to an average of  five sets of measurements. For each measurement, the correlations are calculated from 500 PSD data. The error bar shows the standard deviation.}
	\label{nonlinearity}
\end{figure}

In this experiment,  a  1064nm laser tightly focused by a high NA objective lens (NA=0.85) is used to trap and detect the nanodumbbell in vacuum.  The power of the trapping beam at the focus is about 200 mW. In order to confine the orientation of the nanodumbbell, the trapping laser is linearly polarized along the $x$-axis (Fig.\ref{scheme}). 
The potential energy of a small nanodumbbell in a linearly-polarized Gaussian optical tweezers can be approximately written as
 \begin{align}
U(\alpha,\beta,x,y,z) &= -\frac{1}{4}[\alpha_{\bot}+(\alpha_{\Vert}-\alpha_{\bot})\cos^{2}(\alpha)\sin^{2}(\beta)] \nonumber \\
 &\times \frac{E^2_0}{1+(\frac{z}{z_0})^2} \exp[-\frac{2x^2}{w^2_x(z)}-\frac{2y^2}{w^2_y(z)}],
\label{eq1}
\end{align} 
where $\alpha_{\Vert}$ ($\alpha_{\bot}$) is the polarizability parallel (perpendicular) to the long axis of the nanodumbbell and $E_{0}$ is the electric field amplitude of the trapping laser at the focus.  
$w_{x,y}(z)=w_{x,y}(z=0) \cdot \sqrt{1+z^2/z^2_0}$ is the beam waist radius, and $z_0$ is the Rayleigh range of the optical tweezers. The nanodumbell is therefore trapped close to the focus and simultaneously aligned parallel to the polarization axis of the trapping laser. The equilibrium orientations are $\alpha=0$, $\beta=\frac{\pi}{2}$. Note the trapping potential $U(\alpha,\beta,x,y,z)$ is independent of $\gamma$ because of the rotational symmetry of the nanodumbbell (Fig.\ref{scheme}(c)). The trapping potential as a function of the motion along the $y$ axis $U_y \equiv U(\alpha=0,\beta=\frac{\pi}{2},x=0,y,z=0)$ and the trapping potential as function of the rotation along $\alpha$ direction $U_{\alpha}\equiv U(\alpha,\beta=\frac{\pi}{2},x=0,y=0,z=0)$ are shown in Fig.\ref{scheme}(b).

The out-going 1064 nm laser beam is collected with a collimation lens after the trap. It is sent to four balanced photodetectors to detect the translational and torsional vibrations.  Except the measurement of the torsional vibration in the $\beta$ direction, the optical configurations to detect the motion of the nanodumbbell are similar to those in  previous reports \cite{Hoang2016,Ahn2018}. As shown in Fig.\ref{scheme}(d),
 the signal obtained from the detector monitoring the translational motion in the $x$ direction contains the information of the torsional motion in the $\beta$ direction. The reason is that the rotation of the nanodumbbell along $\beta$ direction will deflect the laser beam in $\beta$ direction, causing a shift of the laser beam along $x$ axis at the position of the detector \cite{Seberson2019}. Because the c.m. motion and the torsional vibration have different frequencies, we can separate them 
with band-pass filters. In this work, we use two different sizes of nanodumbbells. Nanodumbbells consisting of 170 nm-diameter nanospheres are used to study the nonlinear torsional dynamics. Nanodumbbells consisting of two 120 nm-diameter spheres are used for the cooling experiment. Larger nanodumbbells have larger signals, but are more difficult to  trap in vacuum with a 1064 nm laser.

From the measured power spectral densities (PSDs) of a levitated nanodumbbell (Fig.\ref{scheme} (d)), we immediately notice that the torsional peaks ($\alpha$ and $\beta$) around $350$ kHz are much broader than their expected linear linewidths in vacuum. Thus we need to investigate the nonlinear characteristics of the torsional motion. The c.m. trapping potential of optical tweezers can be approximated by a Gaussian potential, which is  not harmonic. This causes nonlinear translational motion of a trapped nanoparticle in vacuum \cite{Gieseler2013,yoneda2017}. For torsional vibrations, the nonlinearity is much stronger because the potential $U(\alpha,\beta,x,y,z)$ is sinusoidally dependent on $\alpha$ and $\beta$ (Fig.\ref{scheme}(b)). The trapping depth for the orientation confinement is also much smaller than that for the c.m. confinements (Fig.\ref{scheme}(b)). To  simplify the problem, we apply feedback cooling to c.m. motions so the nonlinearity caused by the thermal c.m. motion can be neglected. 
We introduce the deviation angles from the equilibrium orientations as  $\xi=\alpha$, $\eta=\frac{\pi}{2}-\beta$. If the deviation angles are small, we can approximate the system as a Duffing nonlinear oscillator. The nonlinear c.m. motion frequencies ($\omega_{x_i}$, $x_i=x,y,z$) and torsional frequencies ($\omega_{\xi}, \omega_{\eta}$)  can be obtained  from Eq. \ref{eq1}:
\begin{equation} \label{eq2}
 \begin{aligned}
\omega^2_{x_i}&=\omega^2_{x_i,0} [1-\frac{\alpha_{\Vert}-\alpha_{\bot}}{\alpha_{\Vert}} (\xi^2+\eta^2)],  \\
\omega^2_{\xi}&=\omega^2_{\xi,0} (1-\frac{2}{3} \xi^2-\eta^2), \\
\omega^2_{\eta}&=\omega^2_{\eta,0} (1- \xi^2-\frac{2}{3}\eta^2).
\end{aligned} 
\end{equation}
Here $\omega_{x_i,0}$, $\omega_{\xi,0}$, $\omega_{\eta,0}$  are intrinsic trapping frequencies when the vibration amplitudes are 0. Based on Eq. \ref{eq2}, a finite-amplitude vibration in any direction will decrease the frequencies in all modes simultaneously.  Thus we expect the frequency fluctuations of all modes to be positively correlated. 

 The measured frequency fluctuations are presented in Fig. \ref{nonlinearity}. 
When the pressure is relatively high (5 Torr), frequency fluctuations of the two modes are small because of the high damping rate (Fig.\ref{nonlinearity} (c)). When the pressure decreases, however, the rarefied surrounding gas is unable to provide enough damping during the measurement time (2 ms). As shown in Fig.\ref{nonlinearity} (a), the thermal motion then causes large frequency fluctuations because of the nonlinearity.  Besides the large frequency fluctuations, another consequence of nonlinearity is the strong correlation of frequency fluctuations in different modes.  As expected from Eq. \ref{eq2}, the relative fluctuations in c.m. frequencies $\omega_{x_i}/\langle\omega_{x_i}\rangle$ at $3\times10^{-3}$ torr are positively correlated (Fig. \ref{nonlinearity}(a)). The normalized correlation of the c.m. frequency fluctuations $\omega_x \star \omega_y$ becomes close to one ($0.93 \pm 0.02$ at $3\times10^{-3}$ torr) as the pressure decreases (Fig. \ref{nonlinearity}(d)), which is similar to the case for a levitated single nanosphere \cite{Gieseler2013}. 
 To further test Eq. \ref{eq2}, we introduce $\omega_{t} \equiv \sqrt{(\omega^2_{\xi}+\omega^2_{\eta})/2}$, and its adjusted relative fluctuation $r_t=1+\frac{6}{5}\frac{\alpha_{\Vert}-\alpha_{\bot}}{\alpha_{\Vert}} (\frac{\omega_t}{\langle \omega_t \rangle}-1)$. For a nanodumbbell with aspect ratio of 1.9, we have $\frac{\alpha_{\Vert}-\alpha_{\bot}}{\alpha_{\Vert}}=0.126$ \citep{Ahn2018}. The adjusted relative fluctuation $r_t$ in torsional frequencies is plotted together with fluctuations in c.m. frequencies in Fig. \ref{nonlinearity}(a). $r_t$  overlaps with $\omega_{x_i}/\langle\omega_{x_i}\rangle$ very well, which agrees with Eq. \ref{eq2}. 
 
\begin{figure}[t]
	\includegraphics[scale=1]{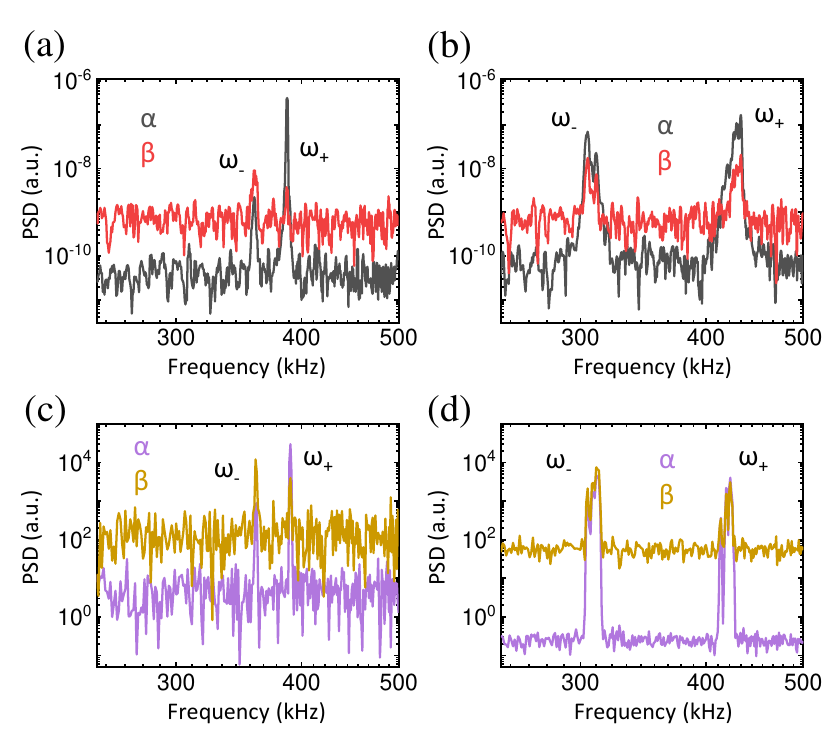}
	\caption{(a), (b): Measured PSDs of motions along $\alpha$ (grey) and $\beta$ (red) directions when $\omega_{c}$ is small (a) or large (b). The PSDs are taken at $3\times10^{-3}$ Torr and the measurement time is 2 ms.  (c), (d): Corresponding simulation results for $\alpha$ (purple) and $\beta$ (yellow) motions when $\omega_{c}/2\pi$ is 17 kHz (c) or 115 kHz (d). The simulated motion time is 1 ms. Random background noise is added to mimic the experiment.}
	\label{precession}
\end{figure}

 However, Eq. \ref{eq2} could not explain all features in the measured torsional PSDs. For example, the measured torsional PSDs have two peaks for both $\alpha$ and $\beta$ motions (Fig. \ref{scheme}(d)). The frequencies of these two peaks ($\omega_+$, $\omega_-$) are negatively correlated at low pressure (Fig. \ref{nonlinearity}(b)), which could not be explained by Eq. \ref{eq2}.
This disagreement is because we have not considered the free rotation of the nanodummbell around its long axis. This rotation will couple the two torsional modes \cite{Seberson2019}: $\ddot{\xi}=-\omega^{2}_{\xi} \xi-\omega_{c}\dot{\eta}$, 
$\ddot{\eta}=-\omega^{2}_{\eta} \eta+\omega_{c}\dot{\xi}$.
Here $\omega_{c}=(I_{z}/I_{x})\omega_{\gamma}$ and $\omega_{\gamma}$ is the angular frequency of its spin around its symmetric axis. Because of the rotation coupling, the solutions for $\xi$ and $\eta$ have two normal modes $\omega_{+}$ and $\omega_{-}$ which can be understood as clockwise and counterclockwise precession modes. They are hybrid modes of the torsional motions. 
We have
\begin{equation} \label{eq5}
\omega_{\pm} =\frac{1}{\sqrt{2}}\left[2\omega^2_t+\omega^2_{c}\pm\sqrt{4\omega^2_t\omega^2_c+\omega^4_c+\Delta^4}\right]^{\frac{1}{2}},
\end{equation}
where  $\Delta^2 = \omega^2_{\xi}-\omega^2_{\eta}$. If $\Delta^4<<4\omega^4_t$, which is the case in our experiment, we have $\omega_t = \sqrt{\omega_+ \omega_-}$. This equation is used to calculate $\omega_t$ shown in  Fig.\ref{nonlinearity}. $\omega_{+}$ and $\omega_{-}$ change oppositely when  $\omega_c$ changes. Due to the Brownian rotation of the nanodumbbell, the two torsional peaks are enforced to move in the opposite direction as can be seen in Fig.\ref{nonlinearity} (b). This competes with the nonlinear  effect (Eq. \ref{eq2}) and eventually becomes superior as the pressure goes down. As a result, the correlation between the frequencies of these two hybrid modes $\omega_{+} \star \omega_{-}$ is negative in high vacuum (Fig.\ref{nonlinearity} (d)).

\begin{figure}[tb!]
	\includegraphics[scale=1]{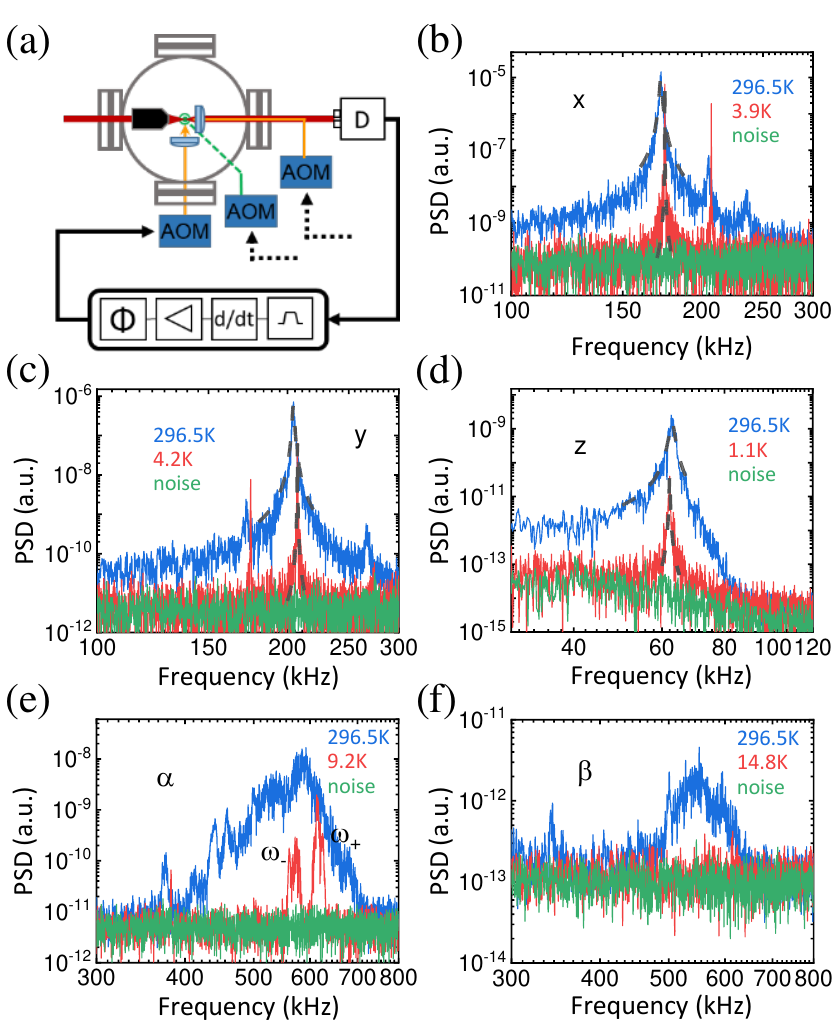}
	\caption{5D cooling of a levitated nanodumbbell. (a) Simplified schematic of the setup showing the sequence of cooling for a single direction. In the real experiment, signals from the particle are collected with four  balanced detectors (D) and processed with five home-built circuits to generate the corresponding feedback signals. The signal obtained from the balanced detector for both $x$ and $\beta$ motions is split to two and fed into two different derivative circuits in order to generate the cooling signals for the motion in $x$  and $\beta$ directions. After processing, signals corresponding to $\alpha$ and $\beta$ motions are added to the signals for $z$ and $y$ motions, respectively (Fig.\ref{scheme} (a)). These signals are then used to modulate the cooling lasers. The results of 5D feedback cooling are shown  in (b) x, (c) y, (d) z, (e) $\alpha$, and (f) $\beta$. The blue curves are PSDs before cooling, and the red curves are PSDs with cooling. The green spectra show the noise levels when there is no particle. The pressure is 1 Torr for blue curves (no cooling), and is $1.8\times10^{-3}$ Torr for red curves (with cooling). The effective temperatures for $x$, $y$ and $z$ motions are calculated based on Lorentzian fittings (shown in grey dashed lines). The effective temperatures for $\alpha$ and $\beta$ motions are calculated by comparing the areas below the PSDs.}
	\label{cooling}
\end{figure}

To further investigate the nature of complex rigid-body motion of a levitated nanodumbbell, we perform numerical simulations of the 6D Brownian motion of a nanodumbbell. Examples of simulation results and experimental results of the PSDs of torsional vibrations are shown in Fig.\ref{precession}. As shown in the PSD plot, the frequency difference of the two hybrid modes $\omega_{+}$ and $\omega_{-}$ becomes smaller when the coupling frequency ($\omega_{c}$) decreases.  Note that $\omega_{c}$ is determined by the geometry of the particle ($I_{z}$/$I_{x}$) and Brownian rotation frequency $\omega_{\gamma}$. This observation means that even though the spin of the nanodumbbell around its symmetric axis does not directly interact with the trapping laser, its angular velocity ($\omega_{\gamma}$) can be monitored by tracking the frequency separation of the two hybrid modes. This result paves a way towards getting access to this ``invisible'' degree of freedom that does not couple to the optical tweezer directly.

To cool the translational and torsional vibrations of a nanodumbbell, we apply three linearly polarized cooling lasers along $x$, $y$ and $z$ directions as illustrated in Fig.\ref{scheme}(a). The wavelength of the $x$ cooling laser is 532 nm, and the wavelengths of the $y$ and $z$ cooling lasers are 976 nm. To avoid interference between the $y$ and $z$ cooling lasers, the 976 nm laser has a short coherent length. The intensities of the $x$ and $y$ cooling lasers are both roughly 1 mW/$\mu$m$^2$, and the intensity of the $z$ cooling laser is roughly 5 mW/$\mu$m$^2$. The scattering forces from the 3 cooling lasers are used to cool the nanodumbbell's c.m. motions.  To cool the torsional vibrations, we intentionally tilt the polarization axes of the $y$ and $z$ cooling lasers by about 10 degrees with respect to the direction of polarization of the trapping laser. Thus the $z$ cooling laser can exert a torque on the nanodumbell to cool its $\alpha$ torsional mode, and the $y$ cooling laser can exert a torque  to cool its $\beta$ torsional vibration mode. The polarization direction of the $x$ cooling laser is kept to be parallel to the $y$ axis and it is not used to cool any torsional degree of freedom. 

As shown in Fig. \ref{cooling}(a), signals about the translational and torsional vibrations of the nanodumbbell  are sent to electronic circuits to control the powers of the three cooling lasers with acousto-optic modulators (AOMs). We use five home-built circuits with bandpass filters, differentiators, and variable gain amplifiers to process the signals for cooling. We use a bandpass filter with a frequency range of 220kHz-5MHz to obtain the $\beta$ signal from the output of the $x$ detector (Fig.\ref{scheme}(d)). The differentiators calculate the derivatives of the motional signals and provide velocity information for cooling. The two torsional cooling signals for $\alpha$ and $\beta$ motions are added on top of the translational cooling signals for $z$ and $y$ motions using adder circuits before feeding into the AOM drivers. The powers of the cooling lasers are modulated as $\Delta P_x=-C_x \frac{d x}{d t}$, $\Delta P_y=-C_y \frac{d y}{d t}-C_{\beta} \frac{d \beta}{d t}$, and $\Delta P_z=-C_z \frac{d z}{d t}-C_{\alpha} \frac{d \alpha}{d t}$ to achieve 5D cooling. Here $C_x$, $C_y$, $C_z$, $C_{\alpha}$, and $C_{\beta}$ are modulation coefficients controlled by variable gain amplifiers.

The results of 5D cooling of a levitated nanodumbbell are shown in Fig. \ref{cooling}. Changes of the PSDs due to feedback cooling are plotted for each degree of freedom. The c.m. motions of the nanodumbbell are cooled to a few K at $1.8 \times 10^{-3}$~torr in all three directions. The energy of the two torsional modes are also reduced by feedback cooling. Since both hybrid modes ($\omega_+$, $\omega_-$) contribute to  the torsional motions along $\alpha$ and $\beta$ directions, we consider both peaks together to extract the motional temperature. The lowest effective temperatures achieved for the two torsional DOFs are 9.2 K ($\alpha$) and 14.8 K ($\beta$) respectively. This is mainly limited by the low signal-to-noise ratio of the $\beta$ signal obtained from the $x$ detector. In fact, the PSD of the cooled $\beta$ vibration is close to its noise level (Fig. \ref{cooling}(f)). In the future, we can add another laser along the $y$ axis to detect this mode more efficiently. 
To our best knowledge, this is the first  report on 5D cooling of a levitated nanoparticle. The nanodumbbell has  six motional degrees of freedom in total. The uncooled degree of freedom, which is the spin motion ($\gamma$), does not directly interact with the laser because of the symmetry of the nanodumbbell. 

In conclusion, we investigate the nonlinear dynamics of a levitated nanodumbbell and demonstrate 5D cooling of its motion.  The frequency fluctuations of the torsional motions are observed to be much  larger than those of the c.m. motions. In the case of the torsional motions, it turns out that the two peaks in the frequency domain are influenced by two distinct factors: nonlinearity and rotation coupling.  The large nonlinearity of the torsional motion could be advantageous for creating nonclassical states and sensing \cite{badzey2005coherent,qvarfort2018gravimetry,RevModPhys.86.1391}. We also demonstrate 5D cooling of a levitated nanodumbbell by developing an  active force and torque feedback cooling method.  The only uncooled degree of freedom is the spin around its long axis which has no direct interaction with the trapping laser. We could successively observe the effect of this rotational degree of freedom via the relative frequency difference of the two hybrid modes of the torsional motions. Thus, the angular frequency of the nanodumbbell's rotation can be observed even if the rotational motion itself is not observable.  This work helps us to better understand  the dynamics of a levitated nonspherical particle and cool its motion to the ground state in all degrees of freedom in future.  Our work is also relevant to cooling of other nonsphereric particles, such as nanorods \cite{Stickler2018probing},  microdisks \cite{SinghPhysRevLett.105.213602,Geraci2013} and mirrors \cite{Guccione2013mirror} for exploring new physics.

\vspace{0.3cm}

We are grateful to supports from the Office of Naval Research under grant No.
N00014-18-1-2371 and the NSF under grant No. PHY-1555035.

%\bibliography{TorsionalCooling_ver_revised}

%

\end{document}